\documentclass[sigconf]{acmart}

\AtBeginDocument{%
  }

\setcopyright{acmlicensed}
\copyrightyear{2018}
\acmYear{2018}
\acmDOI{XXXXXXX.XXXXXXX}
\acmConference[SBES '25]{Brazilian Symposium on Software Engineering}{September 22–26, 2025}{Recife, PE}
\acmISBN{978-1-4503-XXXX-X/18/06}

\PassOptionsToPackage{table,xcdraw}{xcolor}
\usepackage{colortbl}
\usepackage{cleveref}
\usepackage{fontawesome5}
\usepackage{listings}
\usepackage{parcolumns}
\usepackage{color} 
 
\usepackage{amssymb}
\usepackage{booktabs} 
\usepackage{graphicx}  
\usepackage{array}     
\usepackage{multirow}  
\usepackage{tikz,lipsum,lmodern}
\usepackage[most]{tcolorbox}
\usepackage{tabularx} 

\setcopyright{none} 
\renewcommand\footnotetextcopyrightpermission[1]{} 

\newcommand{\kl}[1]{\color{blue} #1 \normalcolor}

\definecolor{codegreen}{rgb}{0,0.6,0}
\definecolor{codegray}{rgb}{0.5,0.5,0.5}
\definecolor{codepurple}{rgb}{0.58,0,0.82}
\definecolor{backcolour}{rgb}{0.95,0.95,0.92}

\lstdefinestyle{mystyle}{
    backgroundcolor=\color{backcolour},   
    commentstyle=\color{codegreen},
    keywordstyle=\color{magenta},
    numberstyle=\tiny\color{codegray},
    stringstyle=\color{codepurple},
    basicstyle=\ttfamily\footnotesize,
    breakatwhitespace=false,         
    breaklines=true,                 
    captionpos=b,                    
    keepspaces=true,                 
    numbers=left,                    
    numbersep=5pt,                  
    showspaces=false,                
    showstringspaces=false,
    showtabs=false,                  
    tabsize=2
}
\newcommand{\phiDetectPerc}{97\%{}}

\newcommand{\gemmaDetectPerc}{91\%{}}

\newcommand{\llamaDetectPerc}{90\%{}}

\begin{document}

\title{Investigating the Performance of Small Language Models in Detecting Test Smells in Manual Test Cases}


\author{Keila Lucas}
 \orcid{0000-0003-0730-5846}
 \affiliation{%
   \normalsize \institution{Federal University of Campina Grande} \country{Brazil}
 }
 \email{keila.santos@copin.ufcg.edu.br}
 
 \author{Rohit Gheyi}
 \orcid{0000-0002-5562-4449}
 \affiliation{%
   \normalsize \institution{Federal University of Campina Grande} \country{Brazil}
 }
 \email{rohit@dsc.ufcg.edu.br}

  \author{Márcio Ribeiro}
 \orcid{0000-0002-4293-4261}
 \affiliation{%
   \normalsize \institution{Federal University of Alagoas} \country{Brazil}
 }
 \email{marcio@ic.ufal.br}

  \author{Fabio Palomba}
 \orcid{0000-0001-9337-5116}
 \affiliation{%
   \normalsize \institution{University of Salerno} \country{Italy}
 }
 \email{fpalomba@unisa.it}

   \author{Luana Martins}
 \orcid{0000-0001-6340-7615}
 \affiliation{%
   \normalsize \institution{University of Salerno} \country{Italy}
 }
 \email{lalmeidamartins@unisa.it}
 
 \author{Elvys Soares}
\orcid{0000-0001-7593-0147}
\affiliation{
  \department{Federal Institute of Alagoas} 
  \country{Brazil}
}
\email{elvys.soares@ifal.edu.br}

\begin{abstract}
Manual testing, in which testers follow natural language instructions to validate system behavior, remains crucial for uncovering issues not easily captured by automation. However, these test cases often suffer from test smells, quality issues such as ambiguity, redundancy, or missing checks that reduce test reliability and maintainability. While detection tools exist, they typically require manual rule definition and lack scalability. This study investigates the potential of Small Language Models (SLMs) for automatically detecting test smells. We evaluate Gemma3, Llama3.2, and Phi-4 on 143 real-world Ubuntu test cases, covering seven types of test smells. Phi-4 achieved the best results, reaching a \(pass@2\) of 97\% in detecting sentences with test smells, while Gemma3 and Llama3.2 reached approximately 91\%. Beyond detection, SLMs autonomously explained issues and suggested improvements, even without explicit prompt instructions. They enabled low-cost, concept-driven identification of diverse test smells without relying on extensive rule definitions or syntactic analysis. These findings highlight the potential of SLMs as efficient tools that preserve data privacy and can improve test quality in real-world scenarios.
\end{abstract}

\keywords{Test Smells; Small Language Models; Manual Testing.}

\maketitle

\section{Introduction}
\label{sec:introduction}

Manual testing is a widely used software verification technique in which developers execute predefined test steps to detect defects in the system~\cite{hauptmann2016,SoaresAORGSMSFB23}. These steps are typically written in natural language and provide instructions on how to interact with the system and assess its behavior. Manual testing remains essential for identifying issues that are difficult to detect using automated tools~\cite{Hauptmann2013}.

Despite its importance, manual testing often suffers from quality and reliability issues. Test descriptions, written in natural language, are frequently created outside formal software engineering practices~\cite{Hauptmann2013}. As a result, they may contain test smells—issues that make test cases harder to maintain, ambiguous, redundant, overly complex or insufficient for proper verification~\cite{Hauptmann2013,juhnke2021clustering,SoaresAORGSMSFB23,junior2020survey, junior2021}. 
A common example of a test smell is the \textit{Ambiguous Test}, in which vague instructions are given, such as \textit{``Select \textbf{a keyboard layout} and click Continue''}. There is no specification of which layout should be selected -- the tester might assume that any layout is acceptable, that the default layout should be chosen, or that a specific layout is implicitly required -- which leads to multiple interpretations.

Improving the clarity and specificity of such instructions is a complex task. Furthermore, existing tools aimed at detecting these smells in natural language have limited scope and require significant implementation effort, often relying on manually defined rules or extensive keyword specifications. Recent studies~\cite{SoaresAORGSMSFB23, peixoto2024effectiveness} have evaluated issues in manual testing, such as ambiguous descriptions, excessive complexity, actions without checks, or inadequate checks, all of which can severely hinder the effectiveness of the testing process. To improve the quality of the testing process, it is essential to detect problems that affect the understanding of the checks, which can negatively impact the maintainability of the software. \looseness=-1

In light of the evolution of foundation models, Small Language Models (SLMs) have been gaining prominence in Natural Language Processing tasks~\cite{van2024survey, lu2024small}. SLMs offer a lightweight, cost-effective, and customizable alternative to Large Language Models (LLMs), which, while powerful, are resource-intensive and less practical for specialized tasks. Moreover, SLMs can be fine-tuned locally, enhancing data privacy and control.
From a testing perspective, SLMs offer a promising and accessible solution for automating the detection of test smells in manual test descriptions. However, it remains unclear to what extent SLMs can contribute to the identification of test smells. 

In this study, we explore the use of SLMs to detect seven types of test smells in manual test descriptions~\cite{Aranda-ease2024}, aiming to improve the quality and effectiveness of manual testing processes through a more accessible and efficient solution. We evaluate Gemma3 (4.3B), Llama3.2 (3.21B), and Phi-4 (14.7B) models with respect to their ability to detect issues in natural language. Our dataset consists of 261 sentences extracted from 143 manual test cases of the Ubuntu operating system. The performance of each model is assessed based on accuracy and \(pass@2\) for identifying each type of smell, focusing on ambiguity, incomplete checks, unnecessary complexity, and redundancy.
The results demonstrate that SLMs are capable of effectively detecting test smells in manual test descriptions. Among the evaluated models, Phi-4 achieved the best performance, identifying 216 sentences with test smells, exhibiting the lowest performance variation across different temperatures, and reaching a \(pass@2\) of \phiDetectPerc{}. In contrast, the Gemma3 and Llama3.2 models performed significantly worse, with \(pass@2\) scores of \gemmaDetectPerc{} and \llamaDetectPerc{}, respectively.

The results show that SLMs can effectively detect test smells in manual test descriptions and, in some cases, even suggest refactorings without explicit instructions. Their use enables easier detection of various types of smells without extensive syntactic rule definitions, while also offering low computational cost and preserving data privacy. Expanding detection to other types of smells -- even in different languages -- can be easily done by incorporating their conceptual definitions. SLMs are a promising solution for improving manual testing and demonstrate practical value in real-world scenarios. \looseness=-1

\section{Research Methodology}
\label{sec:methodology}

The \textbf{\textit{goal}} of this study is to evaluate the performance of the models Gemma3, Llama3.2, and Phi-4 in detecting test smells in manual test cases, from the perspective of researchers, and considering smell types previously cataloged in the literature. We investigate the following \textbf{\textit{research questions}} (\textbf{RQ}s):
\begin{itemize}

    \item[\textbf{RQ$_{1}$}] To what extent can Gemma3 detect natural language test smells?
    \item[\textbf{RQ$_{2}$}] To what extent can Llama3.2 detect natural language test smells?
    \item[\textbf{RQ$_{3}$}] To what extent can Phi-4 detect natural language test smells?
\end{itemize}

To answer our research questions, we selected test case sentences from the Ubuntu manual, each annotated with one of the seven known smells in natural language tests. We applied Directional Stimulus Prompting (DSP)~\cite{li2023guidingllm} to the SLMs and evaluated each model's responses in terms of correctness and coherence using the \textit{pass@k} and semantic consistency metrics.


\smallskip
\noindent
\textbf{\textit{Models selection.}} We selected Llama3.2 because it is more compact than the latest version, Llama3.3. The 3B model is lightweight, and effective in multilingual text generation~\cite{metallama3}. The Gemma3 and Phi-4 models are recent models that achieve robust performance relative to their size, especially in reasoning-focused benchmarks~\cite{team2025gemma,abdin2024phi}. The Phi-4 model provides high performance in NLP tasks, with the advantage of being easily adaptable and implementable on local devices~\cite{abdin2024phi}.
\looseness=-1

\smallskip
\noindent
\textbf{\textit{Model setup.}} The SLMs selected for the study, the Gemma3, Llama3.2, and Phi-4 models, were executed using the Ollama framework~\cite{Ollama2025}, with all default parameters initialized, except for temperature. The temperature is a hyperparameter that controls the randomness of the model's output~\cite{peeperkorn2024,ackley1985}, we evaluated the detection accuracy of the models across 11 temperature settings, which were adjusted within the range from 0.0 to 1.0 (with an increase of 0.1). 

\smallskip
\noindent
\textbf{\textit{Dataset selection.}}
We selected 261 sentences in natural language extracted from 143 manual test cases of the Ubuntu Operating System ~\cite{manual_test_alchemist,ubuntu_manual_tests}. We chose manual test cases from Ubuntu due to its popularity and the large number of open tests that validate various functionalities and behaviors of the system. Each sentence has been manually curated by Aranda \emph{et al.}~\cite{Aranda-ease2024} and represents a set of system verifications and expected results to guide the tester during the execution of the tests. The reference catalog defined by Aranda \emph{et al.}~\cite{Aranda-ease2024} presents seven types of test smells in natural language, as presented in Table~\ref{tab:nl-smell-sample}.


\begin{table}[]
    \centering
    \footnotesize
    \caption{Sample occurrences per test smell.}
    \label{tab:nl-smell-sample}
    \begin{tabular}{lr}
        \toprule
        \rowcolor{gray!30}
        \textbf{Test Smell} & \textbf{Sample} \\
        \midrule
        Ambiguous Test & 90 \\
        Conditional Test & 9 \\
        Eager Action & 69 \\
        Misplaced Action & 8 \\
        Misplaced Precondition & 5 \\
        Misplaced Verification & 16 \\
        Unverified Action & 64 \\
        \bottomrule
        \textbf{TOTAL} & \textbf{261} \\
        \bottomrule
    \end{tabular}
\end{table}

\begin{figure}[H]
\includegraphics[width=0.72\linewidth]{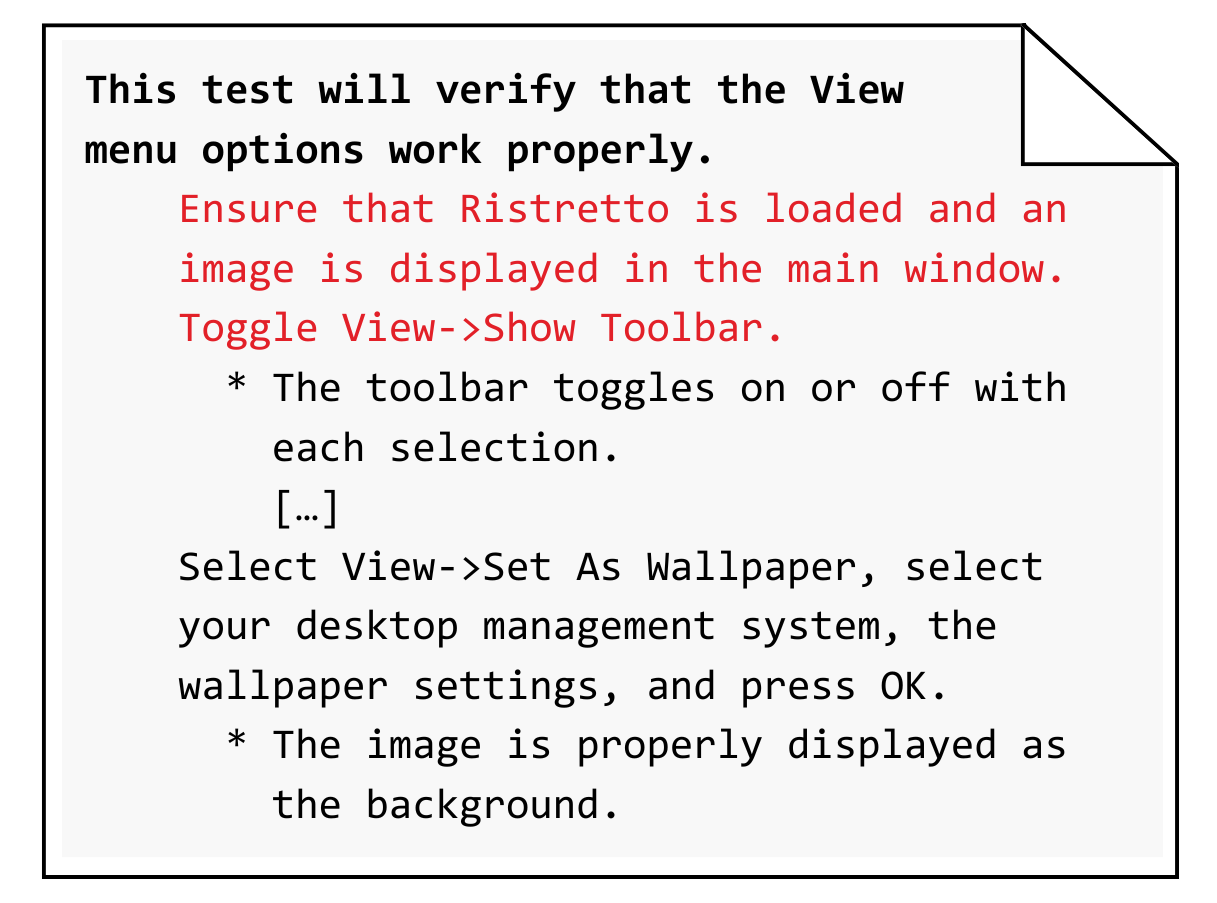}
\vspace{-12pt}
\caption{Ubuntu test case \textit{1600\_Ristretto.}}
\label{fig:test-case-example}
\end{figure}

In the test case \textit{1600\_Ristretto} (Figure ~\ref{fig:test-case-example}), we highlight one example sentence that contains the \textit{Eager Action} test smell, as multiple actions are combined into a single instruction without ensuring that each action is completed before proceeding to the next. If Ristretto is not loaded or an image is not displayed, attempting to toggle the toolbar may fail, and the test might not correctly identify the cause of the error.\looseness=-1

\begin{figure}[]
\includegraphics[width=0.95\linewidth]{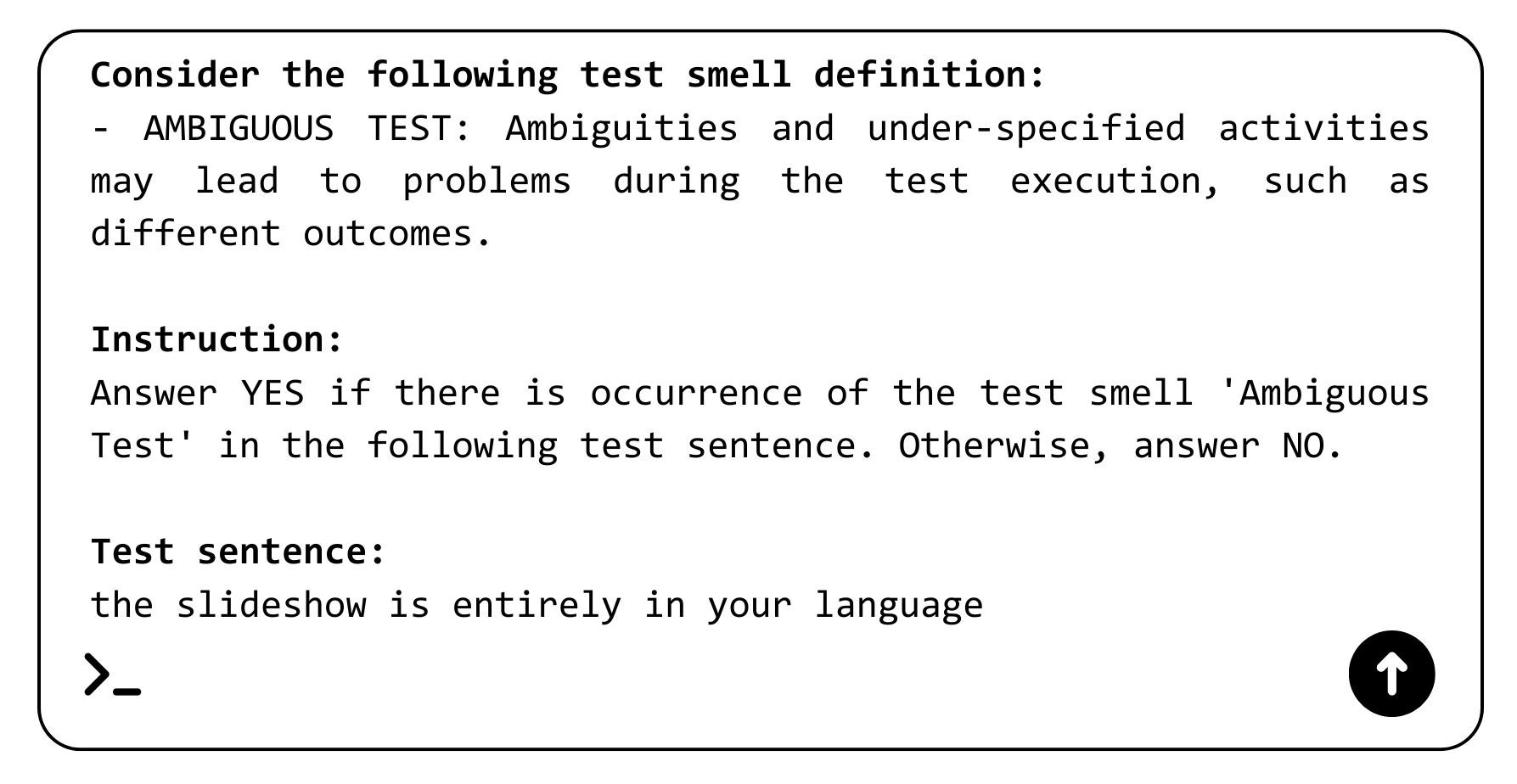}
\vspace{-10pt}
\caption{The Directional Stimulus Prompting.}
\label{fig:dsp-prompting}
\end{figure}

\smallskip
\noindent
\textbf{\textit{Prompt design.}} The queries to the models were guided by DSP. It is a prompting technique designed to optimize the LLM's response in generating the desired outcome. In this approach, we provide a stimulus or clue related to the specific concept of test smell that we aim to detect in the test sentences. In DSP, we structure the prompt by presenting the concept of the test smell, a guiding instruction, and the test sentence for analysis. For each type of test smell, we adjust the concept and the question, as shown in Figure~\ref{fig:dsp-prompting}. To guide the models in detecting test smells, we included the conceptual definition of each type in the prompt, in English, as presented in Table~\ref{tab:definitions}. All responses generated by the models are available in the study’s artifacts~\cite{artefatos}.

\begin{table}[t]
    \centering
    \footnotesize 
    \caption{Test Smell Definitions.}
    \label{tab:definitions}
    \begin{tabular}{m{2.5cm} p{5.2cm}}  
        \toprule
        \rowcolor{gray!30}
        \textbf{Test Smell} & \textbf{Definition} \\
        \hline
        Ambiguous Test         & The smell occurs when ambiguities and under-specified activities may lead to problems during the test execution, such as different outcomes. \\
        \hline
        Conditional Test       & The smell occurs when conditionals introduce uncertainty and variability to the test.\\
        \hline
        Eager Action           & The smell occurs when a single step groups multiple actions. \\
        \hline
        Misplaced Action       & The smell occurs when an action is placed in the verification list.\\
        \hline
        Misplaced Precondition & This smell occurs when a precondition is placed as the first action of the test. \\
        \hline
        Misplaced Verification & The smell occurs when there is a verification in the actions list. \\
        \hline
        Unverified Action      & The smell occurs when there is no verification for a given action. \\
        \bottomrule
    \end{tabular}
\end{table}

The models were implemented and executed via Google Colab, using a T4 GPU and Python. The dataset of sentences was loaded from a CSV file with the help of the Pandas library. All queries to the SLMs were performed in March 2025, and the models were downloaded between March 11\textsuperscript{th} and 14\textsuperscript{th}, as detailed in Table~\ref{tab:model-setup}.

\begin{table}[h!]
\centering
\footnotesize
\caption{SLMs Setup.}
\label{tab:model-setup}
\begin{tabular}{cccc}
\toprule
\rowcolor{gray!30}
& \textbf{Gemma3:4B} & \textbf{Llama3.2:3B} & \textbf{Phi-4:14B} \\
\midrule
Arch        & gemma3      & llama       & phi3 \\
Parameters  & 4.3B        & 3.21B       & 14.7B \\
Quantization & Q4\_K\_M     & Q4\_K\_M     & Q4\_K\_M \\
Updated     & Mar/2025    & Out/2024    & Jan/2025 \\
Download    & Mar/14/2025 & Mar/11/2025 & Mar/13/2025 \\
\bottomrule
\end{tabular}
\end{table}

\begin{figure}[H]
\centering
\includegraphics[width=0.46\textwidth]{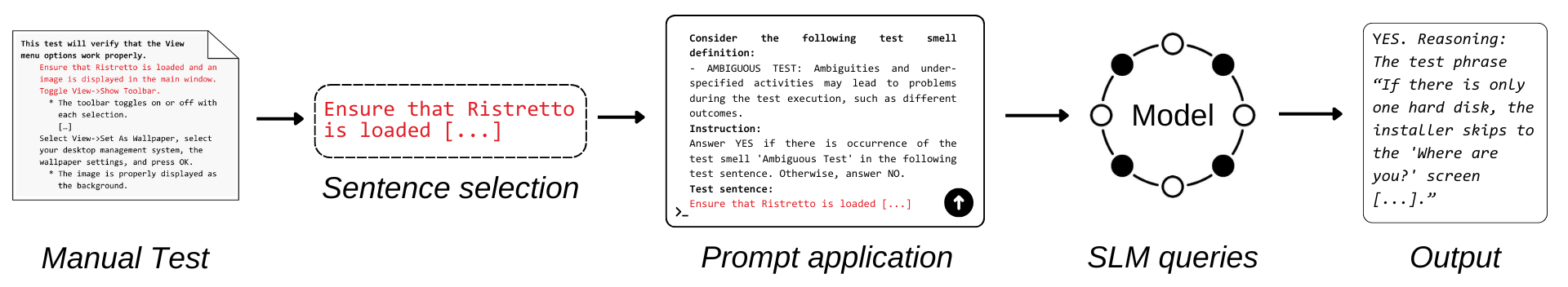} 
\caption{Approach to Manual Test Smell Detection.}
\label{fig:SLM-approach}
\end{figure}

\smallskip
\noindent
\textbf{\textit{Test smell detection workflow.}} The test smell detection process was developed in four phases, as illustrated in Figure~\ref{fig:SLM-approach}. First, we selected 261 natural language sentences extracted from manual tests of the Ubuntu operating system, which were manually reviewed~\cite{Aranda-ease2024}. Next, the test sentences were submitted to the SLMs using DSP. Each model processed the prompt and generated a textual response indicating whether or not it detected the specific test smell in the given sentence. The models justified their answers based on the definition provided in the prompt. For each sentence, we collected 11 outputs (one for each adjusted temperature setting), across two execution attempts.

In the response analysis phase, we evaluated the performance of each SLM using the \textit{pass@k} and \textit{semantic consistency} metrics, aiming to assess both the correctness and coherence of the generated solutions. Responses were considered correct when the model’s justification matched the test smell annotation in the manual baseline. Based on the initial analysis, we identified an opportunity to improve the prompting technique to obtain more accurate, and reliable responses. This enhancement process is discussed in Section~\ref{sec:discussion-meta}. Finally, we conducted a final review, considering the effectiveness of the changes made to the prompt formulation.

\smallskip
\noindent
\textbf{\textit{Pass@k and Semantic Consistency metrics.}} We evaluated the performance of the models by observing the \(pass@k\) metric, which is an evaluation used to measure the probability that, by generating \(k\) samples, at least one of them is correct~\cite{chen2021evaluating}. This metric provides the probability of success when generating multiple samples, reflecting the model's ability to generate correct solutions within a limited number of attempts. 
\section{Analysis of the Results}
\label{sec:results}

Phi-4 model achieved a \(pass@2\) of \phiDetectPerc{}, while the Gemma3 and Llama3.2 models reached a \(pass@2\) of \gemmaDetectPerc{} and \llamaDetectPerc{}, respectively. Regarding the outputs generated by the models, we observed Semantic Consistency (SC), which evaluates how reliably a model reproduces correct outputs in repeated submissions~\cite{ouyang2025}. As shown in Table~\ref{tab:models_pass@k}, all models achieved SC above 65\%, with Phi-4 standing out with high consistency (90\%). Therefore, the model’s reasoning is stable across repeated attempts, surpassing random correctness.

\begin{table}[h!]
\centering
\footnotesize
\caption{Evaluation of SLMs across pass@k, Test Smells, and
Semantic Consistency.}
\label{tab:models_pass@k}
\begin{tabular}{lrrr}
\toprule
\rowcolor{gray!30} & \textbf{Gemma3} & \textbf{Llama3.2} & \textbf{Phi-4} \\
\midrule
\rowcolor{gray!20} \textbf{Accuracy (pass@2)} & 91.1\% & 91.0\% & 97.0\% \\
\midrule
Ambiguous Test         & 73.3\% & 77.0\% & 93.3\% \\
Conditional Test       & 55.5\% & 100.0\% & 78.0\% \\
Eager Action           & 97.1\% & 68.1\% & 85.5\% \\
Misplaced Action       & 68.1\% & 62.5\% & 62.5\% \\
Misplaced Precondition & 100.0\% & 100.0\% & 20.0\% \\
Misplaced Verification & 69.0\% & 75.0\% & 69.0\% \\
Unverified Action      & 39.0\% & 69.0\% & 76.5\% \\
\midrule
\rowcolor{gray!20} \textbf{Semantic Consistency} & 65.0\% & 66.1\% & 90.0\% \\
\bottomrule
\end{tabular}
\end{table}

Despite the lower performance compared to Phi-4, the Llama3.2 model achieved a \(pass@2\) of \llamaDetectPerc{}, with fewer parameters (3B) in its architecture, and being lighter than the Phi-4 model (14B). In scenarios with strict resource constraints, the Llama model can be combined with a second model for test smell detection.

We varied the temperature parameter to assess its impact on test smell detection. Phi-4 consistently outperformed the other models across all settings (77–81\%), with its best result at 0.2. Gemma3 and Llama3.2 showed stable performance overall, peaking at 0.4 and 1.0, with detection rates ranging from 49–54\% and 45–53\%, respectively. 

In terms of accuracy variation across temperature settings, Gemma3 and Phi-4 showed differences of 13 and 12 detections, respectively, between the maximum and minimum accuracy values. Additionally, the Llama3.2 model showed the highest variation, with a 25-point difference in accuracy between temperatures 0.6 and 1.0. This suggests that the temperature increase had a stronger influence on the accurate detection of test smells.

In the overall results, considering the best performance of each model, 253 sentences affected by test smells were detected by at least one of the models. Meanwhile, in 78 sentences, all three models detected the specific test smells. In only 8 sentences of the dataset, none of the models identified the presence of any of the targeted test smells

\section{Discussion \& Further Analysis}
\label{sec:discussion}


\subsection{Explanations}

SLMs generate responses for the detection of test smells, offering detailed explanations for each identified test smell. These justifications, grounded in the models' reasoning, enhances the review process, enabling developers to gain a deeper understanding and identification of the issues. This represents a significant advantage over other detection approaches that focus solely on syntactic analysis. 

In response to our \textbf{RQ$_{1}$}, we observed that some of the Gemma’s responses lacked accompanying explanations, consisted only of binary outputs such as `YES' or `NO'.
The explanations generated by model are generally brief, both for positive and negative results. In the example of the sentence \textit{``The slideshow is entirely in your language''}, the model points out that the sentence is vague: \textit{``It doesn't specify `which' language, or what constitutes \textbf{entirely}''}, but the model's response did not offer deeper insight on the test smell issue, nor does it provide suggestions to adjust the problem. It is essential to evaluate not only the model’s ability to classify sentences as containing test smells or not, but also how it justifies these decisions. The absence of justification limits our ability to assess the reasoning process behind the final decision. \looseness=-1

Regarding the Llama3.2 model (\textbf{RQ$_{2}$}), the model provided explanations for all the evaluated sentences, at each adjusted temperature. In the \textit{Llama3.2 Explanations} box, the negative explanations generated by the Llama model are generally shorter (i.e., contain fewer tokens) than the positive ones. However, some longer explanations were produced, even though they were \textit{False Negatives}.


\begin{tcolorbox}[
  colback=white, coltitle=white, coltext=black, 
  title=Examples of Llama3.2 explanations for Eager Action, center title, fontupper=\scriptsize, fonttitle=\footnotesize,
  left=2pt, right=2pt, top=2pt, bottom=2pt, breakable, enhanced]

\colorbox{gray!15}{\textbf{True Positive}} \\
\textbf{Sentence:} \textit{``Allow the machine to reboot, select the first option at the grub menu''} \\
\textbf{Output:} \textit{Yes. The test sentence can be broken into multiple steps: (1) Allow the machine to reboot, (2) Select the first option at the grub menu. The action of selecting is grouped with rebooting, violating the single-step rule.}

\vspace{4pt}

\colorbox{gray!15}{\textbf{False Positive}} \\
\textbf{Sentence:} \textit{``Select a non-English language and press Enter''} \\
\textbf{Output:} \textit{No. This does not exhibit Eager Action since selecting a non-English language and pressing Enter is a single action group, not multiple. Hence, the answer is NO.}
\end{tcolorbox}

The results of the Phi-4 model confirm the best performance for the task of test smell detection in natural language (\textbf{RQ$_{3}$}). Phi-4 generates standardized explanations, typically showing consistency, whether in positive or negative responses. Even with shorter responses, the model generated explanations for all evaluated sentences across all adjusted temperatures. The model demonstrated significant potential beyond the detection of test smells. It also proposed improvements—even without direct instruction in the prompt—that help developers understand and address test smell issues in test sentences. We highlight one response to an ambiguity issue in the \textit{Example of Phi-4 explanations with DSP} box. 

\begin{tcolorbox}[
  colback=white, coltitle=white, coltext=black, 
  title= Example of Phi-4 explanation with DSP, center title, fontupper=\scriptsize, fonttitle=\footnotesize,
  left=2pt, right=2pt, top=2pt, bottom=2pt, breakable, enhanced]
\textbf{Sentence:} \textit{The machine logins successfully.}

\textbf{Output:} \textit{Indicate which credentials should be used and what the expected outcomes are (e.g., access to the desktop).}
\end{tcolorbox}

\subsection{False Negatives}

All sentences in our dataset are affected by test smells; however, all models produced negative responses for a subset of them. These responses were often accompanied by brief justifications. In some cases, the models identified more general issues which, although potentially relevant, did not align with the specific categories of test smells targeted in our study.

Llama3.2 generated 80 negative responses regarding the presence of test smells in the sentences. We highlight the sentence \textit{``Verify that the full drive space is allocated''} which is recorded in the dataset with the test smell Misplaced Verification, however, the Llama3.2 was unable to identify the issue, providing the following response: \textit{``Since the test sentence only contains one action (`Verify'), and it does not contain any additional actions that would include more than just the `Verify', there is not enough information about other instructions being impacted.''}. 

The Phi-4 model was able to detect the Misplaced Verification smell in the previously highlighted sentence, but it produced \textit{False Negative} responses for this type in five other cases. In the sentence \textit{``note the state of each check mark''}, the model pointed out that \textit{the sentence simply instructs to observe and record the state without verifying it against the expected results}. This reasoning reveals a limitation of the model in identifying test smells when they result in \textit{False Negatives}, especially for those that require context beyond a single sentence, as is the case with the Misplaced Action and Misplaced Verification types.

\vspace{-0.1cm}
\subsection{Limitations on Test Sentences}

The \textit{Misplaced Action} and \textit{Misplaced Verification} smell types require analysis of the affected sentence concerning other correlated sentences within the test instruction. When the model receives an isolated sentence, it becomes challenging to identify an action or verification that is out of sequence. 

In the sentence \textit{``Open some windows''}, the Gemma3 and Phi-4 models indicate that it is a simple action not presented within a list of verifications, as defined by the \textit{Misplaced Action} test smell. The Llama3.2 model states that the test smell is present in the highlighted sentence, but its explanation is inconsistent. It claims that \textit{``there are actions in the verification list (`Open some windows') that are not part of a typical user command or instruction''}, yet it does not specify which verifications are being referenced in its response.

In the responses from Phi-4, we also observed a limitation in identifying the \textit{Misplaced Precondition} type from a single test sentence. For the sentence \textit{``Make sure auto-hide is enabled''}, the model indicated that without additional context to show that this statement is indeed the first action, it is not possible to definitively classify it as an occurrence of the \textit{Misplaced Precondition} smell. In future evaluations, we will assess the deeper detection of these types of smells using a set of sentences or even a complete test case.

\subsection{Can We Do It Better? The Role of Meta Prompting}
\label{sec:discussion-meta}

Given the limitations of the models in detecting test smells, we conducted a deeper analysis using a more refined and targeted prompt. We considered the relevance of applying the Meta Prompting strategy~\cite{hou_metaprompting} to a subset of sentences. This strategy refers to a prompting technique used to guide a language model in generating or refining other prompts.

Based on the configuration defined for the Phi-4 model, we asked it to generate an ideal prompting to detect the test smells \textit{Ambiguous Test}, \textit{Conditional Test}, and \textit{Eager Action}. Only these three types of test smells were evaluated with Meta Prompting, as the Phi-4 model required more time to process the instructions and evaluate the sentences from the \textit{False Negatives} subset. We chose the Phi-4 model due to its superior performance in the test smell detection process, presented in Section~\ref{sec:results}. 

The Meta Prompting design was generated by Phi-4 based on the following query: \textit{I need to create a prompt for scientific research purposes. Specifically, I need a prompt to detect `Ambiguous Test' smells in manual testing. How would you design a prompt for an SLM to perform this task?} 
In response, the model provided a specific prompt design aimed at detecting the \textit{Ambiguous Test} smell. This design included an explanation of the test smell and instructions to guide the detection of issues in test sentences. The query was later adapted to address the \textit{Conditional Test} and \textit{Eager Action}. 

\begin{figure}[H]
\centering
\includegraphics[width=0.8\linewidth]{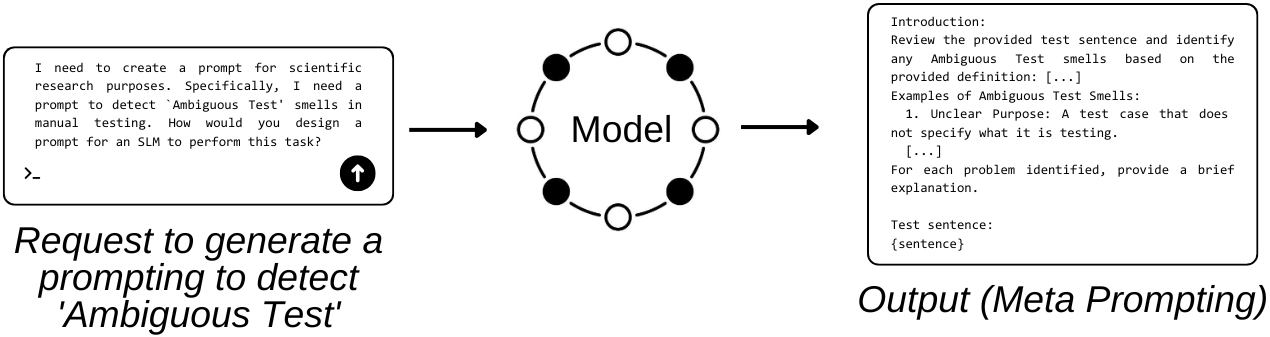}\\
\caption{Improving Prompt Quality for Test Smell Detection Using Meta-Prompting.}
\label{fig:Meta-prompting}
\end{figure}

We evaluated the first output generated by the model for each type of test smell, organized the textual structures, and preserved the order and composition of the instructions created by the Phi-4 model to define the final version of the prompt (see Figure~\ref{fig:Meta-prompting}). 

The instructions generated by the Phi-4 model itself enabled an improvement in the detection of the test smells. For the \textit{Ambiguous Test} type, the prompt was structured to analyze \textit{unclear purpose}, \textit{overly complex logic}, \textit{hardcoded values}, \textit{lack of descriptive naming}, and \textit{redundant assertions}. For the \textit{Conditional Test} type, the instructions were aimed at \textit{verifying tests with deeply nested if-else statements}, \textit{tests relying on specific input values leading to different execution paths}, and \textit{tests using numerous logical operators (AND, OR) in a single condition}. Meanwhile, for the \textit{Eager Action} type, the model was instructed to evaluate \textit{signs of premature execution}, \textit{lack of proper setup}, or \textit{execution of tests without initializing the necessary components}.

We employed Meta Prompting to re-evaluate 19 sentences using the Phi-4 model. The model detected \textit{Ambiguous Test} in six sentences (6/6) and \textit{Eager Action} in 11 sentences (11/11). The re-evaluated sentences were \textit{False Negatives} that the model had previously missed the smells with DSP. We highlight one of the model's responses in the \textit{Example of Phi-4 explanations with Meta Prompting} box. 

\begin{tcolorbox}[
  colback=white, coltitle=white, coltext=black, 
  title= Example of Phi-4 explanations with Meta Prompting, center title, fontupper=\scriptsize, fonttitle=\footnotesize,
  left=2pt, right=2pt, top=2pt, bottom=2pt, breakable, enhanced]
    \textbf{Output:} \textit{The provided test sentence ``The slideshow is entirely in your language'' exhibits several ambiguous test smells based on the definitions provided: \\
    - Unclear Purpose: The test sentence does not specify what aspect of the slideshow it is testing or under what conditions [...].}
\end{tcolorbox}

Regarding \textit{Conditional Test}, the model generated negative responses for the presence of a smell in the sentence \textit{``When ubiquity starts select your language in the left column.''} This sentence is registered in two different test cases: \textit{image/1303\_Live Session} and \textit{image/1497\_Install (live session)}. According to the model: \textit{``The test sentence is simple and clear, lacking any elements that would typically contribute to Conditional Test Logic smells''}. However, the structure \textit{``When [event] [action]''} represents a temporal condition, where selecting the language only occurs after the condition `ubiquity starts' is met. The sentence does not include the common term `if', but there is still a time dependency, as the second event only happens if the first one occurs.

The Meta Prompting strategy enhances the detection of test smells by generating more accurate and detailed explanations for the identified issues. These model-generated explanations play a key role in helping developers understand problems more deeply, facilitating the correction of flaws and the improvement of manual test cases. From the perspective of applying Meta Prompting to the entire dataset, a significant improvement in the performance of the Phi-4 model can be anticipated for the detection of all test smell occurrences.


\subsection{Threats to Validity}

There are some threats to validity that could impact the results and interpretations~\cite{threats-llms-icse-nier-2024}.
We conducted the tests by applying DSP referenced in the literature, manipulating a dataset of natural language sentences that was carefully reviewed by researchers in a previous study. To ensure the validity of this study, the model's responses were reviewed by two authors of the paper to avoid incorrect interpretations in the explanations of the results.

The reproducibility of the results depends on the consistent behavior of SLMs, which may vary with updates or changes in the models. Therefore, we emphasize both the setup of the models and the execution period of the study. 
Additionally, two authors manually reviewed the responses generated by the SLMs to assess potential hallucinations regarding the test smells. This task is challenging, as the models tend to produce lengthy explanations for each evaluated sentence.

The study demonstrates the potential of SLMs in detecting test smells, the conclusions drawn are based on a specific set of natural language sentences. Further testing with a larger dataset of natural language examples is needed to comprehensively validate the findings. The conclusions of the study should be interpreted considering the limitations mentioned earlier and the need for ongoing evaluation and refinement of the capabilities of SLMs.
\vspace{-0.1cm}
\section{Related Work}
\label{sec:related}

Aranda \emph{et al.}~\cite{Aranda-ease2024} proposed a catalog of transformations to automatically remove seven types of test smells from tests written in natural language, such as Ambiguous Test and Eager Action. The authors also developed the Manual Test Alchemist tool, which applies these transformations using NLP techniques. Their evaluation, based on real-world test cases from the Ubuntu OS and a survey with professionals, demonstrated high acceptance (91.4\%) and a promising F-Measure rate of 83.7\%. In the study by Soares et al.~\cite{SoaresAORGSMSFB23}, three systems from different domains were explored, and the authors also highlight promising results from using NLP strategies to reduce the effort required in analyzing tests written in different languages.


Viezaga \emph{et al.}~\cite{veizaga2024} proposed an automated system to detect ambiguous or problematic linguistic patterns in requirements written in natural language. The authors highlight that contextualized rephrasing suggestions can assist analysts in improving requirements quality. 

In the context of test smells in natural language and following the study by Hauptmann \emph{et al.}~\cite{Hauptmann2013}, Rajkovic and Enoiu~\cite{rajkovic2022} developed the NALABS tool to detect smells in requirements and test specifications written in natural language, using keyword lists to assess metrics such as vagueness, referencability, optionality, subjectivity, and fragility. 

Concerning programming languages, Yang \emph{et al.}~\cite{yang2024} addressed the challenges and discussed systematic methods for exploring new types of test smells. In the system proposed by the authors, six detection methods were applied to identify different types of test smells. Similarly, from the perspective of analyzing various types of test smells in programming languages, Lucas \emph{et al.}~\cite{lucas2024} applied the use of LLMs to automatically detect 30 types of test smells. The authors emphasize the potential of language models in identifying test smells, highlighting the models' capabilities to detect and offer improvement suggestions for software tests.

In the context of smell analysis using language models, Wu \emph{et al.}~\cite{wu2024} developed the \textit{iSMELL} system, which integrates LLMs with a set of specialized tools to detect and refactor code smells. The proposed system dynamically selects the most suitable tools to identify three types of complex smells, and the authors emphasize that the tool stands out for its scalability, flexibility, and optimized performance, representing a promising advancement in AI-assisted software engineering. 
\looseness=-1


In our work, we investigate the potential of SLM for automatically detecting test smells, exploring their applicability and effectiveness in identifying issues in natural language test cases. Our goal is to provide a lightweight, scalable, and accessible solution that complements existing methodologies. Phi-4 successfully detects \phiDetectPerc{} of the test smells, demonstrating promising results. The Gemma3 and Llama3.2 models performed less effectively, detecting \gemmaDetectPerc{} and \llamaDetectPerc{} of the smells, respectively. \looseness=-1
\section{Conclusion}
\label{sec:conclusion}

We evaluated the ability of SLMs to detect test smells in natural language test sentences from real-world test cases. Phi-4 outperformed the others, detecting \phiDetectPerc{} of the tests with smells, with a \(pass@2\) of 97\%, while Gemma3 and Llama3.2 detected \gemmaDetectPerc{}. Gemma 4B and LLama 3B detected around 91\% of the cases, which is promising considering they have three to four times fewer parameters than Phi-4 14B. This suggests they may be useful when a lighter and faster model is needed. All data generated by the models is available online~\cite{artefatos}.

SLMs demonstrate strong potential for detecting test smells by providing explanations and suggestions that support refactoring, based only on conceptual definitions of smells and reducing the need for extensive syntactic specifications. This potential is further enhanced by the use of Meta Prompting, which can boost model performance in identifying an even greater number of test smell occurrences in test cases, at a low computational cost.

Future work will extend our approach to detect additional types of test smells and will incorporate agentic methods~\cite{melo2025agenticslmshuntingtest} to assess model performance on larger and more diverse datasets. We are also interested in investigating how foundation models can enable automated refactoring of test cases.

\section*{Acknowledgments}
We thank the reviewers for their feedback. This work was partially supported by CNPq (403719/2024-0, 310313/2022-8, 404825/2023-0, 443393/2023-0, 312195/2021-4), FAPESQ-PB (268/2025).

\bibliographystyle{ACM-Reference-Format}
\bibliography{referencias}

\end{document}